# Relating Band Edge DOS Occupancy Statistics Associated Excited State Electrons Entropy Generation to Free Energy Loss and Intrinsic *V*oc Deficit of Solar Cells

Like Huang[1, a)]

1. Department of Microelectronic Science and Engineering, School of Physical Science and Technology, Ningbo University, Ningbo, Zhejiang 315211, P. R. China;
a) Author to whom correspondence should be addressed: huanglike@nbu.edu.cn

**ABSTRACT**
Ever science the invention of solar cells, thermodynamics has been used to assess their performance limits, guiding advances in materials science and photovoltaic technology to reduce the gap between the practical efficiencies and the thermodynamic limits to photovoltaic energy conversion. By systematically addressing the thermodynamic efficiency losses in current photovoltaic, ultrahigh efficiency photovoltaic can be expected. Currently, the non-radiative recombination of some ultrahigh efficient solar cells is almost completely suppressed, and the radiative recombination loss is then the key to restrict the further improvement of device performance. This work relates the energy band edge electronic density of states (DOS) of semiconductor absorber and transport layer, excited/transfer state electronic entropy to thermodynamically inevitable energy loss during photoelectric conversion in solar cells. On accounts of the basic limitations of thermodynamic laws on the energy conversion process, this work reveals a hidden variable that affects the photovoltaic performance and puts forward the band edge DOS engineering as a new dimension in performance optimization of solar cell apart from the traditional material and defect passivation engineering, etc. This work highlights the great importance of DOS engineering for further improving the performance of any solar cell devices.

## 1. INTRODUCTION

Within just a decade of development, the record photoelectric conversion efficiency (PCE) of perovskite solar cells (PSCs) has reached 25.7%.[1] Apart from the merit of easy fabrication and instructive research experience from traditional solution-processed solar cells, such outstanding and rapid developed efficiency naturally makes people wonder whether halide perovskite (HP) have special properties that make them superior to other materials as photovoltaic absorber.[2] It is commonly believed that HP semiconductors exhibit several remarkable and unique properties which make them ideal for optoelectronics applications: an easily tunable and direct band gap near the optimal gap of Schockley-Queisser limit for single-junction solar cells; a large absorption coefficient, much higher than that of Si and even superior to that of GaAs; sharp absorption edges with a small Urbach energy; large photogenerated carrier lifetime and diffusion length; small and balanced electron and hole effective masses, leading to efficient ambipolar transport with high mobility.[2] Others, however, argue that the mobility of typical HP materials are actually not that high, especially when compared to inorganic semiconductors used as absorbers (GaAs, $\mu_e \approx 8000$ cm$^2$/V s) in high-efficiency PV cells.[3]

As a basic parameter, high open circuit voltage ($V_{oc}$) is of vital significance for high device performance, which demands for low non-radiative recombination rates and come along with high photoluminescence quantum yields (PLQY).[4, 5] For a realistic solar cell with defect induced non-radiative recombination loss, its $V_{oc}$ is directly related to the ability to extract its internal luminescence, as derived by Ross *et al*:[5-8]

$$V_{oc} = V_{oc,rad} + \frac{k_B T}{e} \ln(\eta_{ext}) = V_{oc,rad} - \Delta V_{oc,non-rad} \quad (1)$$

where $V_{oc,rad}$ is the radiative limit of $V_{oc}$, i.e., a thermodynamic limit in the case of 100% radiative recombination. $k_B$ is the Boltzmann constant, $T$ is the temperature, $e$ is the electronic charge, and $\eta_{ext}$ is the external luminescence quantum efficiency. $\eta_{ext}$ is experimentally measurable as PLQY. For any solar cell technology to approach the radiative limit, efficient external electroluminescence (EL) is a necessity. $\Delta V_{oc,non-rad}$ is $V_{oc}$ loss associated with defect induced non-radiative recombination. Obviously, the non-radiative recombination $V_{oc}$ loss is directly related to the external luminescence quantum efficiency $\eta_{ext}$. According to equation (1), by improving the quality of absorption layer and passivating defects to inhibit non-radiative recombination, the device's $V_{oc}$ can gradually approach the radiation limit.[8]

Based on this insightful theory, material/interface engineering and device optimization have greatly reduced the defect state and non-radiation recombination of conventional solar cells and PSCs over the past several decades, making their $V_{oc}$ gradually approach the radiative limit.[9] For example, Grätzel et al. recently reported a record efficiency of 25.7% based on FAPbI$_3$ perovskite with a band gap of 1.53 eV.[1] The $V_{oc}$ of the device is 1.22 V, which is approaching the radiative limit $V_{oc,rad}$ (97.6% of the radiative recombination limit of 1.25 V). Here, the $V_{oc}$ loss ($E_g/e - V_{oc}$) is 0.31 eV, which is very consistent with the common values of about 0.3 V that still exists in all high-efficiency single junction solar cells under the condition of basically eliminating the non-radiative recombination loss, which is, the inevitable radiative recombination $V_{oc}$ loss due to thermodynamics. For such a device, due to the optimized material quality and device design, the space for defect passivation to further improve the $V_{oc}$ is almost zero.

It should be noted that, even with such a low non-radiative recombination loss and such a high $V_{oc}$, this $V_{oc}$ only accounts for 79.7% of $E_g/e$, which means that even if the non-radiative recombination loss is completely eliminated, the $V_{oc}$ of the device still presents remarkable intrinsic radiative recombination loss (about 20% of $E_g/e$). In fact, the dominant energy loss of high-efficiency solar cells, such as GaAs, is also radiative recombination loss. Therefore, as the non-radiative recombination loss in high-efficiency devices is almost eliminated, and the device $V_{oc}$ gradually approaches the radiation recombination limit, new physical insights need to be proposed to further improve the *V*oc and thus device performance.[8]

Recently, the low density of states (DOS) of the valence and conduction band and the associated low effective mass found experimentally and theoretically for HPs have been considered as a possible ingredient to the great success of PSCs, especially for their high *V*oc.[10, 11] However, this result lacks intuitive physical insights. As from the energy dispersion relationship of the band edge, the effective mass of the carrier is positively correlated with the band edge DOS. This means that a small DOS leads to a small effective mass, which usually contributes to a high mobility.[12] From the perspective of semiconductor physics, large mobility is indeed helpful to improve device performance. However, in fact, the mobility of perovskite is not that high.[3] Obviously, the potential contribution of a small DOS to a large *V*oc cannot be understood only from the perspective of semiconductor physics.

In this context, correlating semiconductor physics and thermodynamics may provide new insights, as, although solar cells are semiconductor devices, their ultimate efficiency is essentially limited by thermodynamics. In this work, combined with theoretical derivation and device simulation of PSC, we will point out that the occupation statistics of photogenerated carriers in the band edge electronic density of state leads to entropy generation and thus the reduction of free energy of excited state electrons, which is the essential reason for *V*oc loss of any solar cell.

## 2. RESULTS AND DISCUSSIONS
### 2.1 Theory: Origin of *V*oc from semiconductor device physics

To get a general expression of the *V*oc, let's switching from the above "thermodynamic radiative view" (equation 1) to the "semiconductor picture", and therefore equation 1 can be rewritten using the band gap energy *E*g and charge carrier densities (*n* for electrons and *p* for holes) as will be described by equation (5) below.

According to semiconductor (device) physics, the relationship between *V*oc and the quasi-Fermi levels $E_{Fn}$ and $E_{Fp}$ of electrons and holes in absorber is as follows:[11]

$$eV\text{oc} = E_{Fn} - E_{Fp} \quad (2)$$

Fermi level represents the chemical potential of electrons/holes in semiconductors. Therefore, the increase of chemical potential of electrons and holes after photoexcited electron transition is:

$$\Delta\mu = E_{Fn} - E_{Fp} = \mu_C - \mu_V \quad (3)$$

The electron concentration *n* in conduction band and hole concentration *p* in valence band can be expressed as:[10, 13]

$$n = N_C \exp\frac{\mu_C - E_C}{kT} \quad (4\text{-}1)$$

$$p = N_V \exp\frac{E_V - \mu_V}{kT} \quad (4\text{-}2)$$

To combine formula (2-4), we can get the *V*oc of a solar cell:[10]

$$eV\text{oc} = E\text{g} - k_B T \ln\frac{N_C N_V}{np} \quad (5)$$

wherein,

$$N_{C/V} = 2\left(\frac{2\pi m^*_{e/h} kT}{h^2}\right)^{3/2} \quad (6)$$

is the effective density of states (DOS) of conduction band and valence band, $m^*_{e/h}$ are the effective masses of conduction band electrons and valence band holes, respectively.[13] According to the Pauli exclusion principle, the concentration of conduction band electrons and valence band holes cannot be greater than their corresponding effective DOS, that is $np \leq N_C N_V$, thus:

$$eV\text{oc} = \Delta\mu \leq E\text{g} \quad (7)$$

From equation (5), it can be seen that, for electron/hole pairs excited by photons with energy equal to the band gap, the energy loss caused by factors such as device temperature, conduction band electron and valence band hole concentration and conduction band/valence band effective DOS is the reason why the device *V*oc is less than the band gap *E*g. From the thermodynamic point of view, such a free energy loss of the excited electrons/holes must correspond to a specific kind of entropy increase/generation of the solar cell device as a thermodynamic system.

In thermodynamics, entropy is a fundamental thermodynamic quantity indicative of the accessible degrees of freedom in a system. Entropy increase corresponds to the decrease of the available energy (exergy) of the system. In the Boltzmann formulation, the statistical entropy is:[14]

$$S = k_B \ln \Omega \quad (8)$$

where $k_B$ is the Boltzmann constant and *W* is the number of microstates for a given macrostate. We will prove that the term $k_B T \ln\frac{N_C N_V}{np}$ in equation (5) should have the meaning of entropy. This implies that, although *V*oc scales with *E*g, it cannot reach *E*g for finite temperature, which is due to the second term that can be motivated by the entropy of charge carriers in the accessible density of states in conduction and valence band $N_C N_V$.

## 2.2 Effect of the band-edge DOS of perovskite layer on $V_{oc}$ and device performance from device simulation

According to equation (5), the $V_{oc}$ of a solar cell device depends on the loss term determined by the excitation intensity ($n/p$) and the effective DOS at the conduction band minimum and valence band maximum ($N_C$ and $N_V$) of the absorption layer (which is actually further related to the temperature $T$, which will be discussed later). Next, we will first study the effect of the effective DOS of the absorption layer on the device performance via device simulation with PSCs as an example. The device simulator of the Solar Cell Capacitance Simulator (SCAPS, ver. 3.2.01) developed by University of Gent, was used to model the PSCs.[15]

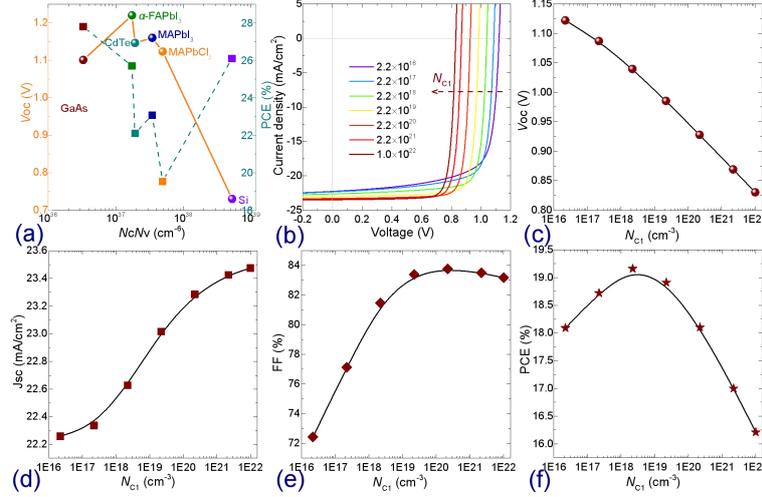

**Figure 1**. $V_{oc}$ of various absorb materials with same electronic parameters except for $N_C$ and $N_V$ (a). Effect of effective density of states $N_C$ at the bottom of conduction band of perovskite on device performance: $J$-$V$ curves (b), $V_{oc}$ (c), $J_{sc}$ (d), FF (e) and PCE (f).

Figure 1a presents the $V_{oc}$ of various absorb materials based solar cells with same electronic parameters except for $N_C N_V$.[16-18] It can be seen that a large $N_C N_V$ value means a low $V_{oc}$, which leads to lower efficiency. Figure 1b shows the variation trend of the $J$-$V$ curves of the device with varying $N_C$ and $N_V$ of the perovskite layer based on device simulation. As can be seen, with the increase of $N_C$ from $1\times10^{16}$ to $1\times10^{22}$ cm$^{-3}$, the $J$-$V$ curves of the device has changed significantly. Specifically, the $V_{oc}$ is reduced from 1.12 to 0.83 V (Figure 1c), which is consistent with the trend of Figure 1a. According to equation (5), the change of $V_{oc}$ with $\ln N_C$ should be linear, while the calculated change trend here deviates slightly from the linear relationship. The specific reasons behind this are not clear here.

Interestingly, the change trend of short-circuit current density ($J_{sc}$) and fill factor (FF) is different from that of $V_{oc}$. As shown in Figure 1d, when $N_C$ increases from $1\times10^{16}$ to $1\times10^{22}$ cm$^{-3}$, the $J_{sc}$ of the device increases slightly from 22.26 to 23.48 mA/cm$^2$. The reason for this result may be that because the absorptivity of the semiconductor is positively correlated with the electronic density of states at the bottom of the conduction band and the top of the valence band, the large $N_C$ leads to the enhancement of absorption and the increase of current density (further explanation will be given below). With the increase of $N_C$ of the perovskite layer, the FF of the device first increases and then saturates (Figure 1e). The specific reason for this obvious change (about 10% absolute change) is not clear. Therefore, the influence of electronic DOS on FF lacks of physical intuition presently. However, the overall result is that with other parameters remain unchanged, the PCE of the device first increases and then decreases with the increase of $N_C$ of perovskite layer, and the performance of the device is the best at $N_C = 1\times10^{18}$ cm$^{-3}$ (Figure 1f). This value is indeed consistent with the DOS of halide perovskite currently reported. Such a small $N_C$ is significantly smaller than that of semiconductor materials such as silicon, which is also the potential reason for the high $V_{oc}$ and high efficiency of PSCs as pointed out in the previous works.[10]

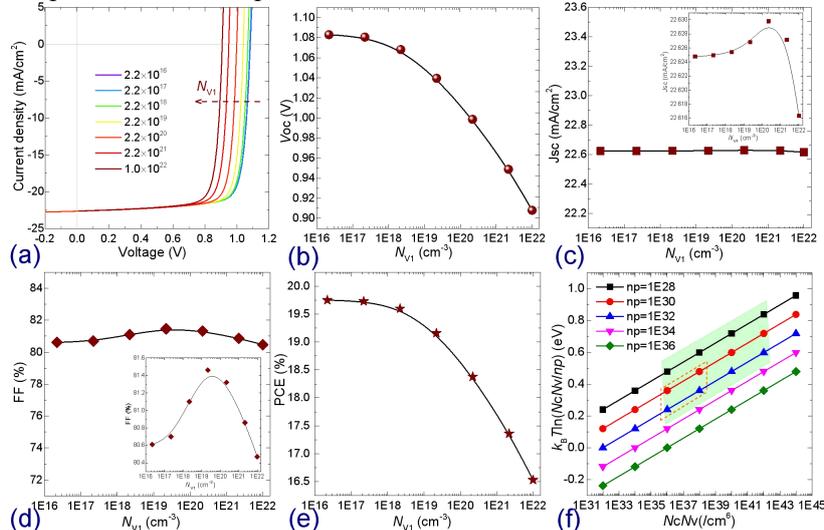

**Figure 2**. Effect of effective DOS $N_V$ at the valence band maximum of perovskite on device performance: $J$-$V$ curves (a), $V_{oc}$

(b), *J*sc (c), FF (d) and PCE (e). *V*oc loss under different excitation intensities and with different $N_C N_V$ values.

Similarly, because of the symmetry of the behavior of electron and hole in semiconductor and p-n junction devices, the valence band maximum effective electronic DOS $N_V$ of the absorption layer may also have a significant impact on the device performance. Figure 2a shows the variation trend of the *J-V* curves of the device with varying $N_V$ of the perovskite layer based on device simulation of the solar cell. It can be seen that with the increase of $N_V$ from $1\times10^{16}$ to $1\times10^{22}$ cm$^{-3}$, the *J-V* curves of the device indeed changed significantly. Just as expected, the *V*oc of the device is reduced from 1.08 to 0.91 V (Figure 2b), which is consistent with the change trend of Figure 1a. Note that, the decay trend here is also nonlinear as mentioned above. While, interestingly, the influence of $N_V$ on the device *J-V* curves is slightly different from that of $N_C$ (mainly reflected in the different influence on the *J*sc). Different from the influence of $N_C$ on the performance of the device, the change of $N_V$ has little influence on the *J*sc and FF of the device (Figure 2c, d). Here, the change trend of FF is difficult to understand due to the lack of direct physical intuition as described above. However, the reason why the change of *J*sc shows such a trend can be attributed to the following reasons: for the fixed $N_C = 22\times10^{18}$ cm$^{-3}$, increasing $N_V$ does not cause a significant increase in absorption, especially when $N_V > N_C$, because the bottom of the conduction band with limited DOS cannot provide more electronic states for electrons excited from valance band to occupy. The overall result of the above changes is that when other parameters remain unchanged, the PCE of the device decreases gradually with the increase of $N_V$ in the absorber layer, especially when $N_V > 1\times10^{18}$ cm$^{-3}$ (Figure 2e).

Figure 2f further shows the dependence of the *V*oc loss of the device on the excitation intensity (*n/p*). It can be seen that under a fixed $N_C/N_V$, the *V*oc loss term decreases and the *V*oc increases with the increase of excitation intensity. Therefore, this is also the basis idea of concentrating photovoltaic (CPV) and the embodiment of its advantages. The green diamond area in Figure 2f shows the numerical range of common solar cells (considering factors such as band gap and light absorption range), while the red dotted diamond area shows the numerical range of PSCs. Note that according to Figure 2f, when *n/p* is about $10^{15}$ - $10^{16}$ cm$^{-3}$ and $N_C/N_V$ is about $10^{18}$ cm$^{-3}$, the *V*oc loss value is very close to the *V*oc deficit (*Eg*/*e*-*V*oc) of about 0.3 eV reported by most efficient PSCs.

Usual irradiation levels (such as AM 1.5 G, 1 sun) are only able to create a low concentration of photogenerated charges (*n*, *p* < $10^{16}$ cm$^{-3}$),[19] whereas the total available states ($N_C/N_V$) of common semiconductor can be about $10^{20}$ cm$^{-3}$, the classic occupancy of electrons/holes is only about 1%. To facilitate discussion, equation (5) can be changed to:

$$eV\text{oc} = E\text{g} + k_B T \ln \frac{np}{N_C N_V} \qquad (9)$$

where $\frac{np}{N_C N_V}$ describes the classic occupation degree of photogenerated nonequilibrium carriers *np* to the effective electron DOS $N_C N_V$. Under general illumination (such as AM 1.5 G, 1 sun), the photogenerated nonequilibrium carriers concentration (*np*: $1\times10^{28}$ - $1\times10^{32}$ cm$^{-3}$) in the absorption layer are much smaller than the value of $N_C N_V$ ($\geq 1\times10^{36}$ cm$^{-3}$), leading to $\frac{np}{N_C N_V}$ far less than 1. Therefore, the second term in equation (9) is negative, and with the increase of the difference between *np* and $N_C N_V$, the lower the occupation, the lower the *V*oc of the device.

The above results imply that for efficient solar cell, the effective electronic DOS at the bottom of the conduction band and the top of the valence band of the absorber layer material should be as small as possible ($\leq 1\times10^{18}$ cm$^{-3}$) and the excitation intensity should be as large as possible to increase the occupancy of photogenerated carriers to the band edge DOS and reduce the *V*oc loss. Halide perovskite materials happen to have a low effective DOS, which may be one of the important reasons for their low *V*oc loss and high device performance, not just the widely considered appropriate band gap, high mobility, long carrier life, high defect tolerance and so on. However, the physical insights behind this are not intuitive presently. Below we will put out new insights to explain this from the perspective of thermodynamics.

## 2.3 Band-edge DOS occupancy statistics, excited state electrons entropy generation and free-energy loss associated *V*oc deficit from thermodynamics

Now, let's turn again to the thermodynamic aspect of solar cells. In fact, with further in-depth consideration, according to equation (5), the *V*oc of a solar cell depends not only on the excitation intensity (proportional to the photogenerated nonequilibrium carrier concentration *np*) and the effective electronic DOS $N_C$ and $N_V$ of the absorption layer, but also on the occupation of the photogenerated nonequilibrium carrier (*np*) to the effective electron DOS $N_C N_V$ and the temperature *T* as a whole. The direct result of this notion is that, for a specific semiconductor absorption layer material, at absolute zero ($T\rightarrow 0$), the maximum exportable *V*oc is its band gap *Eg*/*e*. In fact, since the absolute zero temperature cannot be reached in thermodynamics, the *V*oc of the device will be always less than the band gap *Eg*/*e*, as shown in equation (5), including a negative term. The amount of the negative term depends on the occupation and temperature of the non-equilibrium carrier (*np*) to the effective electronic DOS $N_C N_V$.

According to equation (5) and the above numerical calculation results, the effective electron DOS of the absorption layer significantly affects the performance of the device. Moreover, although equation (5) is changed to (9) and the concept of the classical occupation of photogenerated nonequilibrium carriers on the effective electron DOS $N_C N_V$ is introduced, it still lacks of obvious physical intuition, and the above results are still difficult to understand. The reason may be that we have not considered the quantum properties and quantum identity of electrons, that is electrons are indistinguishable from each other.

In order to reveal the specific mechanism behind the above phenomenon and deeply understand the influence of the effective electron DOS of the absorption layer on the performance of the device, we established relevant physical models and carried out simulation calculation and further analysis below.

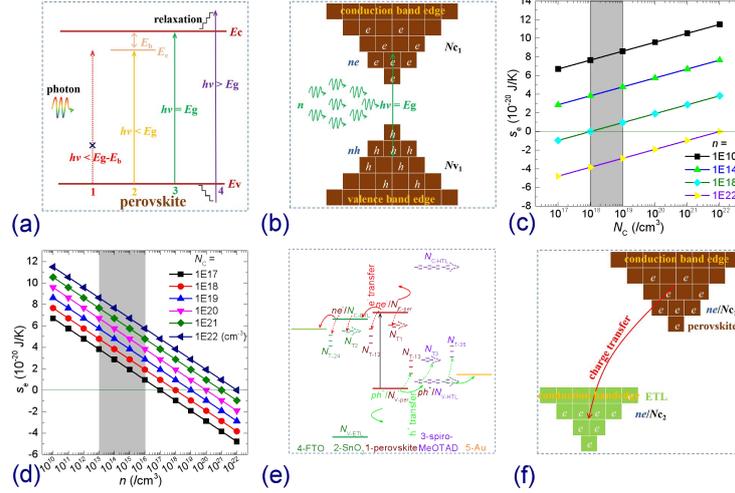

**Figure 3**. Excitation process considered in the perovskite layer (a): 1. Light with an energy below the band gap is not absorbed; 2. Exciton absorption; 3. Light at an energy $hv$ creates an excitation from the valence to the conduction band of a semiconductor an electron-hole pair is formed across the band gap with energy $E_g$; 4. Light at an energy $hv > E_g$ creates an hot electron-hole pair with high kinetic energy. After thermalization in the conduction band an electron-hole pair is formed across the band gap with energy $E_g$ as process 3. Entropy increases due to excited state electrons generation (b). Entropy per charge carrier as a function of the $N_C$ of perovskite (c) and excitation intensity $n$ (cm$^{-3}$, d). Entropy losses due to excited state electrons generation, relaxation, transfer and captured by defect states (e). Entropy increases due to excited state electrons transfer from perovskite to ETL (f).

Considering the quantum properties of electrons and the identity principle, in order to deeply understand the physical meaning behind the above results, we need to use the quantum statistical physics of electrons for further analysis. We first consider the electronic process of the absorption layer under light irradiation. For the optical excitation of the absorption layer, there exist several simultaneous processes as shown in Figure 3a: 1. photons with energy much smaller than the band gap will not be absorbed; 2, photons with energy equal to $E_g$ - $E_b$ will excite excitons (For some semiconductors with significant exciton effects, sub-band gap absorption shown in process 2 can occur, but this is not within the scope of our discussion in this paper.); 3. photons with energy $hv = E_g$ creates a pair of free electron and hole across the band gap; 4. photons with energy greater than or equal to the band gap will excite to produce free electron hole pairs, but the latter will produce hot carriers, and their energy will be quickly transferred to the lattice through relaxation. In order to focus on the essence and key process of the problem, here we only consider the process 3 in which the non-equilibrium free electron/hole pair are generated. Regardless of the hot carrier effect, the available energy (exergy, i.e. the maximum useful work allowed by the law of thermodynamics) $F$ of an excited electron can be expressed as $G = U - TS + PV$. For solar cell as a thermodynamic system, there is no volume change during the process of solar energy conversion. The Gibbs free energy of photogenerated electrons is reduced to Helmholtz free energy:

$$F = U - TS = E_g - TS \quad (10)$$

where $TS$ is the entropy increase associated energy loss of the electrons/holes excited by the photon in the process of photoelectric energy conversion.

Consider equation (5), (10) and Boltzmann's statistical entropy formula (8):

$$S = k_B \ln \Omega \quad (8)$$

according to dimensional analysis and formal analysis, we guess that the term $k_B \ln \frac{N_C N_V}{np}$ in equation (5) (at least in form) has the meaning of entropy! Therefore, the term $k_B T \ln \frac{N_C N_V}{np}$ should have the dimension and physical meaning of energy loss related to entropy change of electron/hole. As such, the term $k_B \ln \frac{N_C N_V}{np}$ should be entropy change of excited electron/hole, and the term $\frac{N_C N_V}{np}$ should be the total number of all possible microscopic states that is occupied by excited electrons/holes in $N_C N_V$.

Next, we will simply prove this understanding with quantum statistical physics analysis of excited electrons and holes. For equation (5), we further expand it as follows:

$$eV_{oc} = E_g - T\left(k_B \ln\frac{N_C}{n} + k_B \ln\frac{N_V}{p}\right) \quad (11)$$

Below, we will see that the last two terms on the right side of equation (11) represent the average entropy change of a single photogenerated electron and hole induced by light excitation, respectively.

Firstly, let us consider the entropy change of (single) photogenerated electron caused by light excitation (Figure 3b). Considering that $n$ photons with energy $hv = E_g$ are absorbed by 1 cm$^3$ semiconductor absorber layer material, and $n/p$ electron/hole pairs will be generated ($n/p$ is the non-equilibrium electron/hole concentration, regardless of recombination, that is EQE = 100%). According to quantum statistical physics, the number of possible microscopic states in which the $n$ electrons occupy $N = N_C$ electronic states at the bottom of conduction band (Figure 3b) is:[13]

$$\Omega = \frac{N!}{(N-n)! \times n!} \quad (12)$$

The spin effect of electrons is not considered here, as one can easily prove that it has little effect on the results follow the above train of thought. According to the statistical significance of Boltzmann entropy, that is, the relationship between entropy and microscopic state number (equation (8)), then entropy change means the change of total microscopic state number: $\Delta S = S' - S = k_B \ln(\Omega_{final}) - k_B \ln(\Omega_{initial}) = k_B \ln(\Omega_{final}/\Omega_{initial})$. Then the entropy change of these excited electrons becomes:

$$\Delta S = k_B \ln \frac{N!}{(N-n)! \times n!} \quad (13)$$

Considering that $n$ (about $10^{14}$ - $10^{18}$ cm$^{-3}$) and $N$ (about $10^{17}$ - $10^{22}$ cm$^{-3}$) are both very large, Stirling approximation:[13]

$$\ln(x!) = x \ln x - x \quad (14)$$

can be used and the above expression approximately equals to:

$$\Delta S = k_B \ln \frac{N!}{(N-n)! \times n!} = k_B \{\ln(N!) - \ln[(N-n)!] - \ln(n!)\}$$
$$\approx k_B \{(N \ln N - N) - [(N-n)\ln(N-n) - (N-n)] - (n \ln n - n)\} \quad (15)$$
$$= k_B \left[N \ln \frac{N}{N-n} + n \ln \frac{N-n}{n}\right]$$

Usually, under the AM1.5 light intensity, $N \gg n$, then $N - n \approx N$, thus the first term of the above formula is approximately zero, and the overall approximation is:

$$\Delta S \approx k_B \left(n \ln \frac{N}{n}\right) = nk_B \left(\ln \frac{N}{n}\right) \quad (16)$$

Let $n = p$, and $N = N_C$, $N = N_V$, the entropy changes of excited $n$ electrons and $p$ holes are:

$$\Delta S_e \approx k_B \left(n \ln \frac{N_C}{n}\right) = nk_B \ln \frac{N_C}{n} \quad (17\text{-}1)$$

$$\Delta S_h \approx k_B \left(p \ln \frac{N_V}{p}\right) = pk_B \ln \frac{N_V}{p} \quad (17\text{-}2)$$

Then the average entropy of a single electron/hole becomes:

$$\Delta s_e = \frac{\Delta S_e}{n} = k_B \ln \frac{N_C}{n} \quad (18\text{-}1)$$

$$\Delta s_h = \frac{\Delta S_h}{p} = k_B \ln \frac{N_V}{p} \quad (18\text{-}2)$$

These are actually the last two terms in the expansion (11) of the $V_{oc}$ formula. Therefore, the last two terms on the right side of equation (11) indeed represent the average entropy change of a single photogenerated electron and hole induced by light excitation, respectively (Figure 3b). We call it excited state electron entropy generation. It shows that the $V_{oc}$ loss term $k_B T \ln \frac{N_C N_V}{np}$ in equation (5) is the sum of free energy loss associated with average entropy changes of a single electron and hole induced by light excitation multiplied by the temperature $T$.

This suggests that the entropy change related energy loss in the $V_{oc}$ formula (5) is related to the excitation intensity (illumination intensity, then $n = p$) and the electron density of states ($N_C N_V$) of the absorption layer as a semiconductor, or to the occupation of the electron state by the photogenerated electron/hole under a certain excitation intensity. That is, under the same excitation light intensity, if $n = p$ is constant, the entropy change of electron/hole increases with the electron DOS, resulting in the increase of energy loss and the decrease of $V_{oc}$. On the other hand, for the same electron DOS, the weaker the light intensity, the smaller the $n/p$, the larger the electron hole entropy and the lower the $V_{oc}$. This suggests that we can increase the $V_{oc}$ by increasing the light intensity (which is a common and practical approach, such as via concentrating photovoltaic) and reducing the "hidden variable" electron DOS of a semiconductor absorb layer (which is not actually reported experimentally). Considering the usual AM1.5 G sunlight with fixed intensity, the absorption layer with lower electron DOS are the potential factors that contributes to a high $V_{oc}$. Therefore, we predict that the band edge DOS engineering will be a potential direction to further increasing the $V_{oc}$ of solar cells in the future. In short, we prove that the negative term in the $V_{oc}$ formula is the energy loss related to the change of excited electron entropy, so it is easy and more intuitively to understand that the $V_{oc}$ of the device changes with DOS.

According to equation (18-1), we calculate the relationship between the average entropy increase of a single excited electron and the effective electron DOS $N_C$ at the bottom of the conduction band of the absorption layer. As shown in Figure 3c and d, in the $N_C$ value range of common semiconductor absorption layer ($1\times10^{18}$ - $1\times10^{20}$ cm$^{-3}$), under the illumination intensity of AM1.5 G ($n$: $1\times10^{14}$ - $1\times10^{16}$ cm$^{-3}$),[19] the average entropy increase of a single excited state electron is about on the order of $1\times10^{-20}$ J/K. The average entropy increase of a single excited hole should have a similar value. Under all excitation intensities, the average entropy increase of a single excited state electron/hole increases with the increase of $N_C/N_V$. Theoretically, when $n = N_C$ and $p = N_V$, the $V_{oc}$ loss of the device is zero, even if the temperature $T$ is not zero. When $n > N_C$, the photogenerated nonequilibrium electrons will preferentially occupy the electronic state at the bottom of the conduction band, and the excess photogenerated nonequilibrium electrons will have to occupy a higher conduction band energy level. This formula will be meaningless. In fact, in order to occupy a higher conduction band level, these excess photogenerated nonequilibrium electrons will have to absorb photons with energy greater than the band gap, which is not within our above hypothesis ($hv = E_g$).

In actual solar cell devices, there will be various defect states at each functional layer and interfaces (Figure 3e). Photogenerated carriers will also transfer to these defect levels and occupy the corresponding interband defective electronic states, which will also cause entropy increase and energy loss (Figure 3e). Since we mainly consider the intrinsic $V_{oc}$ loss of the solar cell related to the band edge DOS occupation, here we do not consider the entropy increase and energy loss related to the defect state.

## 2.4 Effect of DOS on absorption coefficient and *J*sc

As discussed above, we intuitively understand the potential impact of DOS on the $V_{oc}$ of a solar cell based on thermodynamic entropy. On the other hand, the absorption coefficient of a semiconductor material is directly related to its electronic DOS at conduction band and valence band according to the pioneering work by Stephen et al. They showed that the optical absorption coefficient is given by:[20]

$$\alpha(hv) = D^2(hv) \int_{-\infty}^{\infty} N_C(E) N_V(E - hv) dE \quad (19)$$

with

$$J(hv) = \int_{-\infty}^{\infty} N_C(E) N_V(E - hv) dE \quad (20)$$

as the joint density-of-states (JDOS) function and $D^2(hv)$ being the optical transition matrix element. Obviously, the absorption coefficient $\alpha$ is positively correlated with the electronic DOS of conduction band and valence band. This should be the reason why the *J*sc in the above calculation results decreases with the decrease of the electronic density of states (Figure 1b, d and Figure 2a, c), although the reduction is not significant. However, because the increase amplitude of $V_{oc}$ is greater than that of *J*sc, the overall result is that a properly low electronic DOS improves the efficiency of the device (Figure 1f and Figure 2e). This means that the trade-off between the *J*sc and the $V_{oc}$ is not only related through the band gap, but also through the band edge DOS. Therefore, our work here emphasizes the necessity of band edge DOS engineering based on conventional band gap engineering.

## 2.5 Effect of the band-edge DOS of charge transport layer on *V*oc and device performance

By analogy with excited state entropy in absorber layer, we speculate that when excited state electrons/holes are transferred to the corresponding charge transport layer (CTL), entropy increase (we call it transfer state electron entropy generation) will occur again because new occupation of electronic states is involved-excited state electron transfer induces entropy increase (Figure 3f and Figure 5a). Therefore, we calculate the effect of the electronic DOS of the transport layer on the performance of the device. Firstly, we study the effect of the electron DOS of the electron transport layer (ETL) on the performance of the device. The results are shown in Figure 4.

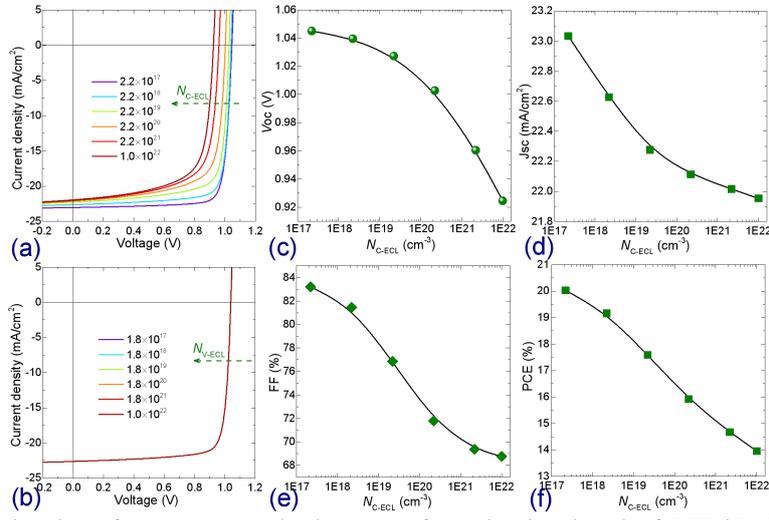

**Figure 4**. Effect of effective density of states $N_C$ at the bottom of conduction band of ETL/ECL on device performance: *J-V* curves (a), *V*oc (b), *J*sc (c), FF (e) and PCE (f). Effect of effective density of states $N_V$ at the valence band maximum of ETL/ECL on device performance: *J-V* curves (b).

Just as expected, DOS of the ETL (dominantly $N_C$) does significantly affect the photovoltaic performance of the device. In particular, its conduction band DOS $N_C$ significantly affects the performance of the device, which implies that when the photogenerated electrons are transferred to the conduction band of the ETL, a new entropy increase will be generated due to the redistribution and occupation of electrons (Figure 5b). However, it is interesting that the change of valence band density of states of ETL has no effect on the performance of the device at all. We continuously changed the order of magnitude and found that all *J-V* curves coincided completely. The *V*oc, *J*sc and FF of the device have no change. Considering the fact that photogenerated electrons are transferred to the conduction band rather than the valence band of the ETL (Figure 5a), this result is very natural and easy to understand. Figure 5 proposes a model to reveal the influence of ETL/ECL and HTL DOS on the entropy increase of excited state electron/hole.

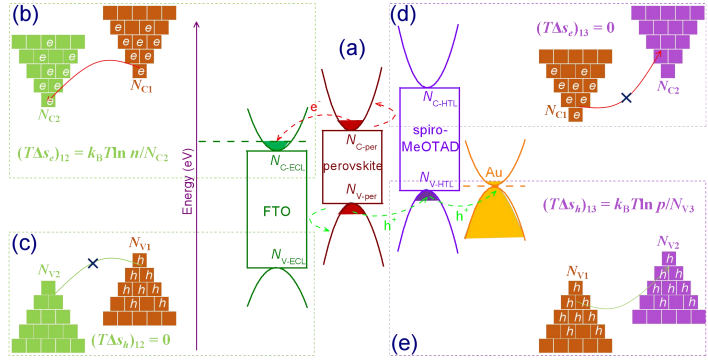

**Figure 5**. Schematic diagram of charge transfer process among energy bands and proposed model for electronic entropy generation of charge transfer states at absorber/CTL interface: (b) conduction band electron entropy produced at the perovskite/ETL interface, (d) zero conduction band electron entropy produced at the perovskite/HTL interface; (c) zero valence band hole entropy produced at the perovskite/ETL interface, (d) valance band hole entropy produced at the perovskite/HTL interface.

Moreover, due to the deep valence band energy level of the ETL, the injection of photogenerated holes in perovskite into its valence band is disadvantageous in energy and is therefore prohibited (Figure 5a). Therefore, the valence band electron state density $N_V$ of the ETL has no contribution to the increase of electron/hole entropy, thus the free energy loss of electron and *V*oc loss of device (Figure 5c). These results suggest that for the ETL/ECL with a main function of collecting electrons and blocking holes, it mainly causes thermodynamically inevitable energy loss and *V*oc reduction of the device through the increase of excited state electron entropy generation. This enlightens us that in order to achieve more efficient (perovskite) solar cells, it is very important to use ETL materials with as small CB effective electron DOS as possible.

In addition, according to the model of Figure 5, if the semiconducting ETL and HTL are replaced by insulators, the relevant entropy increase and energy loss associated with interface charge transfer can be zero, as electrons/holes are transferred by quantum tunneling but not passing through the CB and VB of semiconductor based charge transport layer (ETL and HTL). This may inspire new solar cell device design.

Similarly, according to the reverse symmetry of electron/hole transfer and the reverse symmetry of energy band structure of each functional layer of the whole device, it can be expected that for the hole transport layer (HTL), the change of valence band electron DOS will have a great impact on the photovoltaic performance of the device, because the redistribution of photogenerated holes in VB of HTL from VB of perovskite will produce entropy increase and energy loss during the injection into its valence band (Figure 5e). On the other hand, the bottom level of the conduction band energy of the HTL is shallow, and the photogenerated electrons in perovskite cannot be injected, so the electrons do not produce entropy increase and free energy loss at this interface (Figure 5d). Therefore, the conduction band electron density of states of the HTL has no effect on the performance of the device.

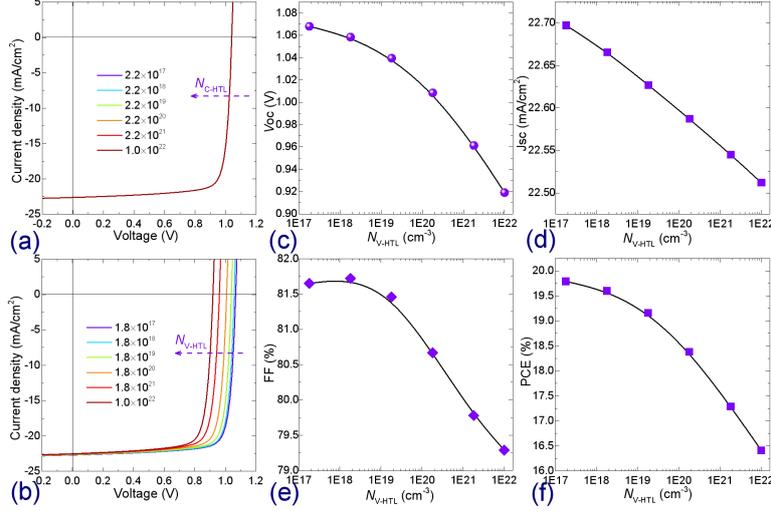

**Figure 6**. Effect of effective density of states $N_C$ at the bottom of conduction band of HTL on device performance: *J-V* curves (a). Effect of effective density of states $N_V$ at the valence band maximum of HTL on device performance: *J-V* curves (b), *V*oc (c), *J*sc (d), FF (e) and PCE (f).

The actual results, as shown in Figure 6, are indeed as expected. As can be seen from Figure 6a, when the effective electron DOS $N_C$ at the bottom of the conduction band of HTL is continuously changed, the *J-V* curve of the device does not change, which is consistent with the expected guess. It can be seen from Figure 6b, but when all other parameters remain unchanged, the *J-V* curve of the device changes significantly with the increase of the effective electron density of states $N_V$ at the top of the valence band. When $N_V$ increases from $1\times10^{17}$ - $1\times10^{22}$ cm$^{-3}$, the *V*oc of the device decreases rapidly from 1.07 to 0.92 V (Figure 6c). According to the above analysis, this is because with the increase of $N_V$ of HTL, the photogenerated holes transferred to the valence band of HTL will produce greater entropy increase, resulting in greater energy loss. Interestingly, as the $N_V$ of HTL increases, the *J*sc of the device decreases slightly (Figure 6d), from 22.67 mA/cm$^2$ to 22.51 mA/cm$^2$. The specific reason behind this subtle change is not clear presently. The FF shows a more complex change behavior (Figure 6e). With the $N_V$ of HTL, FF first increases from 81.65% to 81.72%, and then decreases rapidly to 79.29%. The specific reasons behind this are unclear. As a direct result, with the $N_V$ of HTL, the PCE of the device decreases monotonically from 19.79% to 16.41% (Figure 6f). This enlightens us that in order to achieve more efficient (perovskite) solar cells, it is very important to use HTL materials with as small valence band maximum effective electron DOS as possible. For example, one of the potential reasons why 2D perovskite can improve the performance of devices may be that it is similar to 3D perovskite with low VBM effective electron DOS.

**2.6 Entropy increase and free-energy loss associated with excited state electron transfer**

Accordingly, considering the introduction of CTL and the entropy increase associated with electron transfer due to electron re-occupancy of the DOS of CTL, it is necessary to introduce a new term into formula (2) to change it into:

$$eV\text{oc} = E\text{g} - k_B T \ln \frac{N_C N_V}{np} - k_B T \ln \frac{N_{C\text{-ETL}} N_{V\text{-HTL}}}{np}$$
$$= E\text{g} - k_B T \ln \frac{N_C}{n} - k_B T \ln \frac{N_V}{p} - k_B T \ln \frac{N_{C\text{-ETL}}}{n} - k_B T \ln \frac{N_{V\text{-HTL}}}{p} \quad (21)$$

In the formula, each item is energy loss related to photogenerated electrons induced entropy increase, photogenerated holes induced entropy increase, excited electron transfer entropy increase and excited hole transfer entropy increase.

Considering that the actual device may contain multiple ETL/HTLs, the above formula can be further expressed as:

$$eV\text{oc} = E\text{g} - k_B T \ln \frac{N_C}{n} - k_B T \ln \frac{N_V}{p} - k_B T \sum_i \ln \frac{N_{C\text{-ETL-}i}}{n} - k_B T \sum_i \ln \frac{N_{V\text{-HTL-}i}}{p} \quad (22)$$

According to formula (14) or (14'), for any ETL/HTL based on semiconductor material in the device, if the effective electron state density at the bottom of conduction band (valence band maximum) is large, significant additional entropy related energy loss and voltage loss will be introduced. This also enlightens us that it is also a feasible way to reduce the use of semiconductor based transport layer (simplifying device structure) or using semiconductor materials with lower effective electron density of states at the bottom of conduction band (valence band top) as electron (hole) transport layer. Alternatively, directly using a non-semiconductor material (insulator as mentioned above) as a carrier selective interface layer to replace the semiconductor based CTL may potentially reduce the *V*oc loss. For the former, halide perovskite can be used as an alternative material as CTL to construct simple p-n or conventional p-i-n junction all perovskite solar cells because of its naturally low band edge DOS. For the latter, it is a feasible path to use ultra-thin wide band gap materials to build MIS structured solar cells. The recently reported work with certification efficiency of 25.5% use ultra-thin broadband gap perovskite as the interface layer. Wu et al. used insulating layer to improve the performance of electronic layer free devices to 21%, and the $V_{oc}$ was also significantly improved.

## 2.7 Inspired novel device design for high $V_{oc}$ and high efficiency

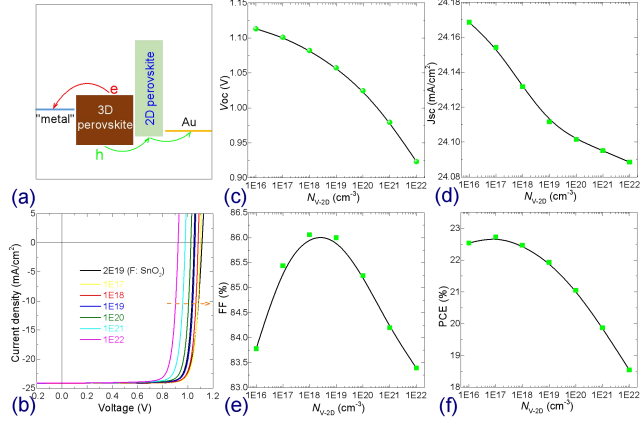

**Figure 7**. 3D-2D mixed dimension all perovskite p-n heterojunction solar cells (a) with 2D perovskite as HTL. Effect of $N_V$ of 2D perovskite HTL on the device performance: (b) $J$-$V$ curves, (c) $V_{oc}$, (d) $J_{sc}$, (e) FF, (f) PCE.

In order to demonstrate the potential beneficial effect of low DOS of the perovskite absorb layer and CTL on the device performance, we further simulate the device performance of 3D/2D mixed dimension all perovskite p-n heterojunction solar cell with respect of the effect of $N_V$ of the 2D perovskite HTL on the device performance. Figure 7a gives the energy band structure of the 3D/2D mixed dimension all perovskite p-n heterojunction solar cell. Figure 7b displays the $J$-$V$ curves of the device with different $N_V$ of the 2D perovskite HTL (with the $N_C$ and $N_V$ of 3D perovskite fixed at 2.2×18 and 1.9×19 /cm$^3$). From Figure 7c, with the increase of $N_V$ of 2D perovskite HTL, $V_{oc}$ decreases significantly. Although the $J_{sc}$ also decreases, the overall margin in value reduction is very small (Figure 7d). The FF shows a bell shaped change with the increase of $N_V$ (Figure 7e). Accordingly, the PCE of the device reaches its maximum value when the $N_V$ of 2D perovskite HTL is about 1×17 /cm$^3$.

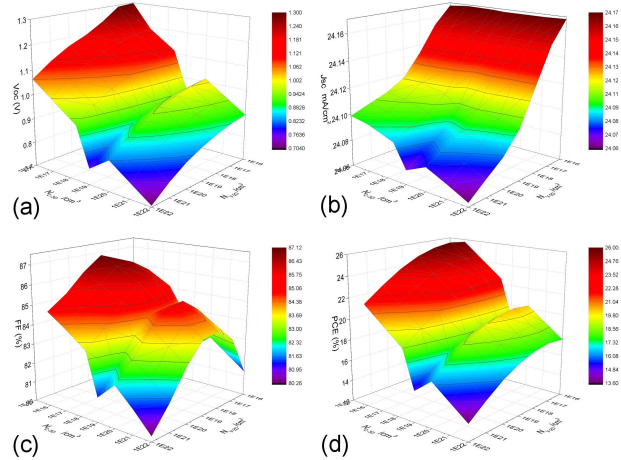

**Figure 8**. The joint effects of band edge DOS of 3D perovskite absorption layer ($N_{C-3D}$) and 2D perovskite HTL ($N_{V-2D}$) on the final device performance: (a) $V_{oc}$, (b) $J_{sc}$, (c) FF, (d) PCE.

In order to further reveal the advantages of all perovskite heterojunction solar cell, we further calculated the joint effects of band edge DOS of 3D perovskite absorption layer and 2D perovskite HTL on the final device performance. From Figure 8a, when the band edge DOS of 3D and 2D perovskite is small simultaneously, the $V_{oc}$ of the device is maximized. While, the $V_{oc}$ is minimum when both DOS are at their maximum. $V_{oc}$ of the device decreases with the increase of DOS of 3D and 2D perovskite. The decrease margin of $J_{sc}$ with both DOS is very small, although it mainly reduces with $N_V$ of 2D perovskite HTL. The range of $N_C$ of 3D perovskite studied seems to have little effect on $J_{sc}$, which may be because the thickness for perovskite is thick enough to ensure sufficient light absorption (Figure 8b). The changes in FF with both DOS are complex and difficult to analyze as there are too many factors that affects FF (Figure 8c). On the whole, because the decrease amplitude of $J_{sc}$ is very small, and both DOS mainly affects the $V_{oc}$ and followed by FF, the variation tendency of PCE with both DOS is almost consistent with that of $V_{oc}$ and FF with both DOS (Figure 8d).

## 2.8 Relationship between band edge DOS and effective electron mass

In fact, for many of the key properties of solar cell absorber material that directly determines the photovoltaic performance, i.e. absorption (absorption coefficient α), charge transport (carrier mobility μ) and recombination, effective electron mass ($m_e^*$) appears in all of their function expressions.

Usually, effective electron mass or effective DOS are features of the band structure that are always calculated in the context of computational material screening efforts. The effective DOS is related to the energy dependent DOS close (within a few $k_BT$) to the band edges. A parabolic approximation to the energy dispersion relation (energy vs. momentum, i.e. $E$-$k$ relation) of the bands is often adopted although many semiconductors have dispersion relations that become non-parabolic away from the conduction band minimum (CBM) or valence band maximum (VBM). For example, the relationship between the energy $E$ and momentum $k$ of an electron near the conduction band edge can be approximated by:[13]

$$E(k) - E_C = \frac{\hbar^2 k^2}{2 m_e^*} \quad (23)$$

where $\hbar = h/2\pi$ is the reduced Planck constant and $m_e^*$ is the effective mass of electron in CB. Analogous equations can be written down for the valence band hole. In this case, the effective mass of the carrier is determined by the second-order derivative of the band edge $E$-$k$ function relationship:[13]

$$\frac{1}{m_{e/h}^*} = \left(\frac{2\pi}{h}\right)^2 \frac{d^2 E}{dk^2} \quad (24)$$

For halide perovskite, due to its highly anisotropy of crystal structure and electronic structure, the curvature of $E(k)$ might be different along different directions in $k$-space thus the $m_e^*$ in general is tensorial. However, we ignore this difference and adopt the approximation of isotropy ($m_e^*$ is the same in all directions of $k$-space), which is reasonable according to the material calculation results. Obviously, the larger the curvature of $E(k)$ near the band edge, the smaller the $m_e^*$ of the carrier. In addition, the $m_e^*$ of carriers also determines the effective electronic DOS at the band edge according to equation (6):[13]

$$N_{C/V} = 2 \left(\frac{2\pi m_{e/h}^* kT}{h^2}\right)^{3/2} \quad (6)$$

Therefore, the carrier effective mass $m_e^*$ is closely related to the band edge energy structure, which correlates the curvature of the band edge energy dispersion $E(k)$ and the effective electron DOS: The smaller the $m_e^*$ of the carrier, the lower the effective electronic DOS at the band edge.

For practical materials, if the anisotropy of lattice and electronic structure is significant, the different orientation of crystal relative to the substrate surface will also lead to the obvious difference of carrier transport performance in absorber material and photovoltaic performance of the device.[21] The crystal orientation and crystal plane dependent photovoltaic performance of PSCs was proposed in our earlier work, which has been confirmed by a large number of subsequent works on the optimization of crystal orientation to improve device performance presently.

According to equation (16), equation (2) can be transformed into:

$$eV_{oc} = E_g - k_B T \ln\left[4\left(\frac{2\pi k}{h^2}\right)^3 T^3 \frac{(m_e^* m_h^*)^{3/2}}{np}\right] \quad (25)$$

As can be seen from equation (25), $V_{oc}$ is a multivariate function of band gap $E_g$, temperature $T$, electron/hole effective mass $m_{e/h}^*$ and excitation intensity $n/p$. Moreover, the specific values of these variables are basically experimentally measurable. This means that $E_g$, $T$, $m_{e/h}^*$ and $n/p$ can be regulated through practical experiments, so as to effectively control the $V_{oc}$, and carefully study the specific influence law of the changes of the above variables on the $V_{oc}$ of the device.

## 2.9 Effect of temperature on $V_{oc}$ and device performance

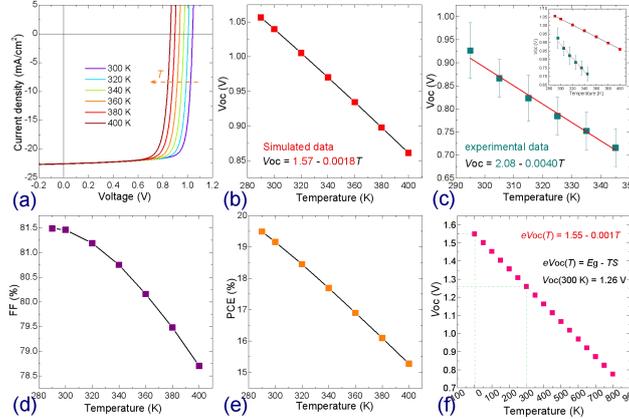

**Figure 9**. Effect of temperature on device performance: $J$-$V$ curves (a), $V_{oc}$ (b), $V_{oc}$ (c, experimental data), FF (d), and PCE (e). Dependence of $V_{oc}$ on temperature $T$ with $S$ almost unchanged (f).

Temperature also affects the entropy related energy loss. In fact, it is precisely because the temperature cannot be zero (the third law of thermodynamics) that the entropy related energy loss cannot be zero. A solar cell dissipates most of the incident solar power as heat, leading to an increase in the cell temperature above the ambient. The amplitude of the temperature increase is largely dependent on the illumination level to which the cell is submitted and the efficiency with which the residual heat is extracted from the cell. The effect of temperature increase on the solar cell performance is studied via device simulation. Figure 9a shows the effect of temperature on the device *J-V* curve from device simulation. The increase of temperature mainly leads to a significant decrease in the *V*oc of the device (Figure 9b), followed by the fill factor (Figure 9d) and little effect on the *J*sc (Figure 9a).

The results of Figure 9 also show that the performance of the device is better at low temperature, and the performance of the device decreases significantly with the increase of temperature (Figure 9b, e). This is because the device has lower entropy related energy loss at low temperature, which is also the potential of perovskite and other solar cells in low-temperature environments such as cold polar regions or space. It is worth noting that Figure 9c shows the change results of the device *V*oc with temperature in the actual experiment. It can be seen that the *V*oc of the actual device decreases faster with the increase of temperature than the result from device simulation (Figure 9c), which may be caused by the reduction of band gap caused by lattice expansion or phase transition of the actual material at high temperature, which is not considered in the device simulation.

For $MAPbI_3$ PSCs, assuming a band gap of 1.55 eV and a maximum *V*oc of 1.26 V at 300 K, thus entropy $S$ can be calculated to be 0.29/300 eV/K ≈ 0.001 eV/K at 300 K. Assuming that entropy $S$ does not change with temperature, the function of the *V*oc with respect to temperature can be expressed as:

$$eV_{oc}(T) = 1.55 - 0.001T \qquad (26)$$

This gives the result that when the temperature tends to be zero, the *V*oc tends to be $E_g/e$ (Figure 9f).

However, the actual situation may be more complicated. According to equation (5), the temperature coefficient of *V*oc is derived as follows:[22]

$$e\frac{dV_{oc}}{dT} = \frac{dE_g}{dT} - k_B T \ln\frac{N_C N_V}{np} + k_B T\left(\frac{1}{n}\frac{dn}{dT} + \frac{1}{p}\frac{dp}{dT}\right) \qquad (27)$$

The third summand in the Eq. has a minor contribution to the derivative because the terms $dn/n$ and $dp/p$ are likely to be very small. The second summand can be expressed in a more useful form:[22]

$$\frac{dV_{oc}}{dT} \approx \frac{d}{dT}\left(\frac{E_g}{e}\right) + \frac{V_{oc} - E_g/e}{T} \qquad (28)$$

From Eq. (28) it is inferred that the temperature dependence is mainly related to the term $(V_{oc} - E_g/e)/T$, which after integration provides a linear relationship $V_{oc}$-$T$ (Eq. 28) as observed experimentally. The main result is that $dV_{oc}/dT$, being the negative temperature coefficients larger in the case of small *V*oc values. It should be noted here that this temperature coefficient is originated by the temperature dependence of the Boltzmann function. In addition, heating reduces the size of the energy gap.

## 3. CONCLUSIONS

In summary, the application of the second law of thermodynamics to semiconductors relates the DOS of semiconductor absorber and transport layer, excited/transfer state electronic entropy to *V*oc loss of solar cell. The results reveal the free energy and intrinsic *V*oc loss of solar cells from the perspective of excited state electron entropy generation. This indicates a feasible direction for further reducing the electron entropy generation through the regulation of electronic density of states, so as to further improve the device *V*oc and efficiency. This further reveals that why perovskite solar cells have high *V*oc and efficiency, as lower effective masses and smaller effective electron density of state leads to smaller entropy generation, and thus smaller free energy loss and *V*oc loss. This work highlights the great importance of band edge DOS engineering for further improving the performance of solar cell devices.

**Appendix** (Another derivation based on Sackur-Tetrode entropy equation)

Another understanding of this point of view can start with the Sackur-Tetrode equation for the entropy of monatomic ideal gases.[23]

Now consider the occupation of electrons over the density of states of semiconductor absorber of a solar cell. For solar cell under dark, the valence band of absorber would be completely occupied and the conduction band would be empty for any temperature *T*. Under this condition, the entropy $S$ of the electrons would be zero, as there is only one possibility of producing a fully occupied valence band and an empty conduction band. While, promoting electrons from the valence band to the conduction band by light excitation would then lead to not only an increase in the energy, but also an increase in the entropy as well. This is because there are now many possibilities for removing an electron from any of about $10^{22}$ states/cm$^3$ of the valence band and exciting it to any of $10^{22}$ states/cm$^3$ of the conduction band.

In 1912, Sackur and Tetrode independently put forward an equation for the absolute entropy of a monatomic classical ideal gas, which is known as the Sackur-Tetrode equation. This is a pioneering investigation about 100 years ago which incorporates quantum considerations. This entropy $S$ at temperature $T$ can be written as:

$$\frac{S}{N} = k_B\left\{\frac{5}{2} + \ln\left[\frac{V}{N}\left(\frac{2\pi m k_B T}{h^2}\right)^{3/2}\right]\right\} = \sigma \qquad (29)$$

where $N$ is the number of particles in the gas, $k_B$ is Boltzmann's constant, $V$ is the volume of the gas, $m$ is the mass of a gas particle and $h$ is Planck's constant. Obviously, $\sigma$ represents the (average) entropy of a single particle in the gas.

In solar cell, light excitation produces electrons in conduction band (CB) and holes in valence band (VB), which relax almost instantaneously (on the submillisecond time scale) to the conduction band minimum (CBM) and the valence band maximum (VBM) and to reach thermal equilibrium with the lattice semiconductor. A large number of such excited electrons form electron gas, which can be seen as ideal monoatomic gas.

Appling the Sackur-Tetrode equation to the electron gas and accounting for the fact that electrons (and holes) have two spin orientations, which doubles the number of possible states, for electrons/holes in a semiconductor, the equation is modified to:

$$\sigma_{e/h} = k_B \left\{ \frac{5}{2} + \ln\left[ \frac{2}{n} \left( \frac{2\pi m^* k_B T}{h^2} \right)^{3/2} \right] \right\} \qquad (30)$$

where $n = N/V$ is the spatial density of electron and $m$ is replaced be $m^*$ of a semiconductor.

According to the equation 6, equation 31 can be re-written as:

$$\sigma_{e/h} = k_B \left( \frac{5}{2} + \ln \frac{N_C(N_V)}{n(p)} \right) \qquad (30')$$

Obviously, the only difference ($5/2 k_B$) of the equation 31' and 19 is the terms on the right. Note that the Boltzmann constant $k_B \approx 1.38 \times 10^{-23}$ J/K, while the calculated average entropy of a single carrier as given in Figure 3c is on the order of $10^{-20}$ J/K, almost a thousand times that of $k_B$. Therefore, in equation 31', compared with the second term on the right, the first term can be ignored. This leads to:

$$\sigma_{e/h} = k_B \left( \frac{5}{2} + \ln \frac{N_C(N_V)}{n(p)} \right) \approx k_B \ln \frac{N_C(N_V)}{n(p)} = \Delta s_{e/h} \qquad (31)$$

This means that, the Sackur-Tetrode entropy of equation 30 and 31 is in fact the average entropy of a single carrier that we derived above, equation 19.

It has been proved that each conduction band electron (hole) in this gas has a mean kinetic energy of $3/2\ k_B T$ and volume energy $k_B T$ ($= pv$) with the potential energy $E_C$ ($-E_V$) as well as an entropy $\sigma_{e/h}$. The free energy or electrochemical potential $\mu_{e/h}$ of an excited state electron (hole) is the sum of these terms minus the entropy $\sigma_{e/h}$ per charge carrier:

$$\mu_e = E_C + \frac{5}{2} k_B T - T\sigma_e = E_C + \frac{5}{2} k_B T - k_B T \left\{ \frac{5}{2} + \ln\left[ \frac{2}{n} \left( \frac{2\pi m_e^* k_B T}{h^2} \right)^{3/2} \right] \right\} \qquad (32\text{-}1)$$

$$\mu_h = -E_V + \frac{5}{2} k_B T - T\sigma_h = -E_V + \frac{5}{2} k_B T - k_B T \left\{ \frac{5}{2} + \ln\left[ \frac{2}{p} \left( \frac{2\pi m_h^* k_B T}{h^2} \right)^{3/2} \right] \right\} \qquad (32\text{-}2)$$

Substituting the effective mass $m^*$ of the electrons (holes) and effective density of states of the energy band from equation (16) for $m$ we find:

$$\mu_e = E_C + \frac{5}{2} k_B T - k_B T \left( \frac{5}{2} + \ln \frac{N_C}{n} \right) \qquad (33\text{-}1)$$

$$\mu_h = -E_V + \frac{5}{2} k_B T - k_B T \left( \frac{5}{2} + \ln \frac{N_V}{p} \right) \qquad (33\text{-}2)$$

Thus, we have:

$$\Delta\mu = E_{Fn} - E_{Fp} = \mu_C - \mu_V = \mu_e + \mu_h = E_g - k_B T \ln \frac{N_C N_V}{np} \qquad (33)$$

This result, obtained directly from entropy analysis of electron gas, is consistent with equation (5).

**ACKNOWLEDGMENTS**
This work was finally supported by the National Natural Science Foundation of China (NFSC) (No. 61904182), and the K. C. Wong Magna Fund in Ningbo University (No. 61904182).

**AUTHOR DECLARATIONS**
**Conflict of Interest**
The authors have no conflicts to disclose.

**DATA AVAILABILITY**
The data that support the findings of this study are available from the corresponding author upon reasonable request.